\documentclass{article} 
\usepackage{epsf}
\usepackage{multicol}

\def\etal{et~al.}
\def\avdelta{\langle \delta \rangle}   
\def\zlim{z_{\rm lim}}                 
\def\lesssim{\mathrel{\hbox{\rlap{\hbox{\lower4pt\hbox{$\sim$}}}\hbox{$<$}}}}
\def\gtrsim{\mathrel{\hbox{\rlap{\hbox{\lower4pt\hbox{$\sim$}}}\hbox{$>$}}}}

\advance \topmargin by -\headheight
\advance \topmargin by -\headsep
\evensidemargin \oddsidemargin
\AtBeginDocument{
		\baselineskip=12pt plus 0.005pt minus 0.005pt}

\newenvironment{references}[0]{
\medskip
\subsection*{References}
\footnotesize \interlinepenalty=10000\hyphenpenalty=10000\parindent=0pt}{}
\def\reference{\par\hangindent=1.2em\hangafter=1}

\makeatletter
\renewcommand\section{\@startsection {section}{1}{\z@}%
                 {-.7\baselineskip \@plus\baselineskip \@minus\baselineskip}%
                                  {0.3\baselineskip}%
                                   {\normalfont\large\bfseries\centering}}
\renewcommand\subsection{\@startsection {subsection}{2}{\z@}%
                                  {-0.7\baselineskip \@minus0.07\baselineskip}%
                                   {0.3\baselineskip}%
                                   {\normalfont\normalsize\bfseries\centering}}
\renewcommand{\@seccntformat}[1]{{\csname the#1\endcsname}\hspace{0.65em}}
\makeatother

\begin{document}
{\noindent \sc\footnotesize Astronomy Letters, in press\\
	\normalfont Translation from Russian, with some modifications}
\medskip
\bigskip

\begin{center}
{\LARGE\bf 
		       The Correlation of Gamma-Ray Bursts

			with Active Galactic Nuclei.

}
\medskip
\bigskip
{\large R.\ A.\ Burenin, A.\ A.\ Vikhlinin, O.\ V.\ Terekhov, 
			and S.\ Yu.\ Sazonov}

\medskip
{\small\it 
	    Space Research Institute, Russian Academy of Sciences

		Profsoyuznaya 84/32, 117810 Moscow, Russia}

\bigskip
\end{center}

\begin{abstract}

We search for angular correlation of gamma-ray bursts with cataloged
quasars, BL Lac objects, and AGN using a large sample of relatively
well-localized bursts detected by WATCH on board GRANAT and EURECA, IPN, and
BATSE (327 bursts total). A statistically significant (99.99\% confidence)
correlation between GRB and $M_B<-21$ AGN in the redshift range $0.1<z<0.32$
is found.  The correlation with AGN is detected, with a lower significance,
in three independent GRB datasets. The correlation amplitude implies that,
depending on the AGN catalog completeness, 10\% to 100\% of bursts with peak
fluxes in the range $3-30\times10^{-6}$~erg~s$^{-1}$~cm$^{-2}$ in the
100--500 keV band are physically related to AGN. The established distance
scale corresponds to the energy release of order $10^{52}$~ergs per burst.

\end{abstract}

\medskip

\begin{multicols}{2}

\section{Introduction}

Gamma-ray bursts are distributed isotropically on the sky; their peak flux
distribution shows the lack of faint bursts compared to the expectation for
a homogeneous distribution in Euclidean space (see e.g.\ Fishman \& Meegan
1995 for review). These observed properties of GRB are reproduced by two
popular models: 1) GRB arise in an extended Galactic halo with a core
radius of $\sim 100$~kpc and 2) GRB are located at cosmological distances.

The Galactic halo models are challenged by the observed isotropy of the
burst positions. The isotropy constraints much improved recently, and as a
result, most variants of the halo model are no longer viable (Briggs et al.\
1996).  An additional argument against the Galactic halo model comes from
{\em Einstein} data. The {\em Einstein} IPC sensitivity is sufficient to
detect bursts from the halos of nearby galaxies, but they were not found in
the data (Hamilton et al.\ 1996).

In the cosmological model, the burst isotropy and departures of $\log N -
\log S$ from the Euclidean $S^{-3/2}$ law are explained naturally.  The
minimum redshift $z=0.835$ of the optical transient associated with
GRB~970528 (Metzger \etal\ 1997) is a decisive evidence in favor of the
cosmological model.  Optical spectroscopy of the GRB counterparts is
complicated by the faintness of associated optical transients and their
relatively featureless spectrum (e.g., no strong emission lines were
observed in the GRB~970528 optical transient).  Therefore, indirect
estimates of the GRB distance scale are still useful.

The shape of the GRB peak flux distribution implies that sources of the
dimmest BATSE bursts are at $z\sim1$, assuming no source evolution (Emslie
\& Horack 1994; Fishman \& Meegan 1995 and references therein). For some
plausible evolution of GRB volume density, the above distance estimate can
vary by a factor of $\gtrsim 2$ (Horack et al.\ 1995). Using a different
approach, Quashnock (1996) derived $z>0.25$ for the dimmest BATSE bursts by
cross-correlating the third BATSE catalog with the known large scale
structure at low redshifts.

Some earlier studies already searched for a direct relationship between GRB
and other astrophysical objects at cosmological distances. Kolatt \& Piran
(1996) and Marani et al.\ (1997) found that GRB from the third BATSE catalog
are correlated with Abell clusters. However, using more precise GRB
localizations, Burenin et al.\ (1997), Hurley et al.\ (1997), and Gorosabel
\& Castro-Tirado (1997) have not found any correlation with Abell clusters.
Furthermore, the amplitude of correlation of well-localized GRB and $z<0.1$
Abell clusters is lower than expected in the case of the same spatial
distribution of GRB and optically luminous matter (Burenin et al.\ 1997).
If this result is interpreted as an indication that bright and
well-localized GRB are at greater distances than nearby Abell clusters, the
dimmest BATSE bursts should be at $z>0.3$, in agreement with Quashnock's
(1996) results.

At still higher redshifts, $z\sim0.1-1$, a natural choice is to search for
correlation of GRB with quasars and other flavors of active galactic
nuclei, which comprise the majority of known objects at these high
redshifts. To search for correlation of AGN and GRB is also attractive
because some theoretical models relate bursts to physical processes in
active nuclei (e.g.\ Lejter 1980, Carter 1992). Some attempts to search for
such a correlation have been undertaken earlier. A marginal evidence for
excess of QSO in the small Interplanetary Network (IPN) GRB error boxes has
been found (Vrba \etal\ 1995). However, Webber \etal\ (1995) and Gorosabel
\etal\ (1995) found that the number of QSO and AGN in GRB error boxes
is consistent with random. These earlier analyses were based on a smaller
number of bursts with good localizations than available at present.  Citing
a recent work, Schartel \etal\ (1997) have found a correlation of
well-localized GRB from the third BATSE catalog with radio-quiet QSO. The
strongest correlation was detected at the $>99.7\%$ confidence level for
intrinsically bright QSO with $z<1$.
	
The goal of this work is to search for a correlation of GRB with QSO and AGN
using all available data for good, with $\lesssim 1^\circ$ uncertainty,
bursts localizations. We use $H_0=50$~km~s$^{-1}$~Mpc$^{-1}$ and $q_0=0$.

\begin{figure*}[htb]
\begin{minipage}[t]{0.47\textwidth}
\epsfysize=1.1\textwidth
\centerline{\epsffile{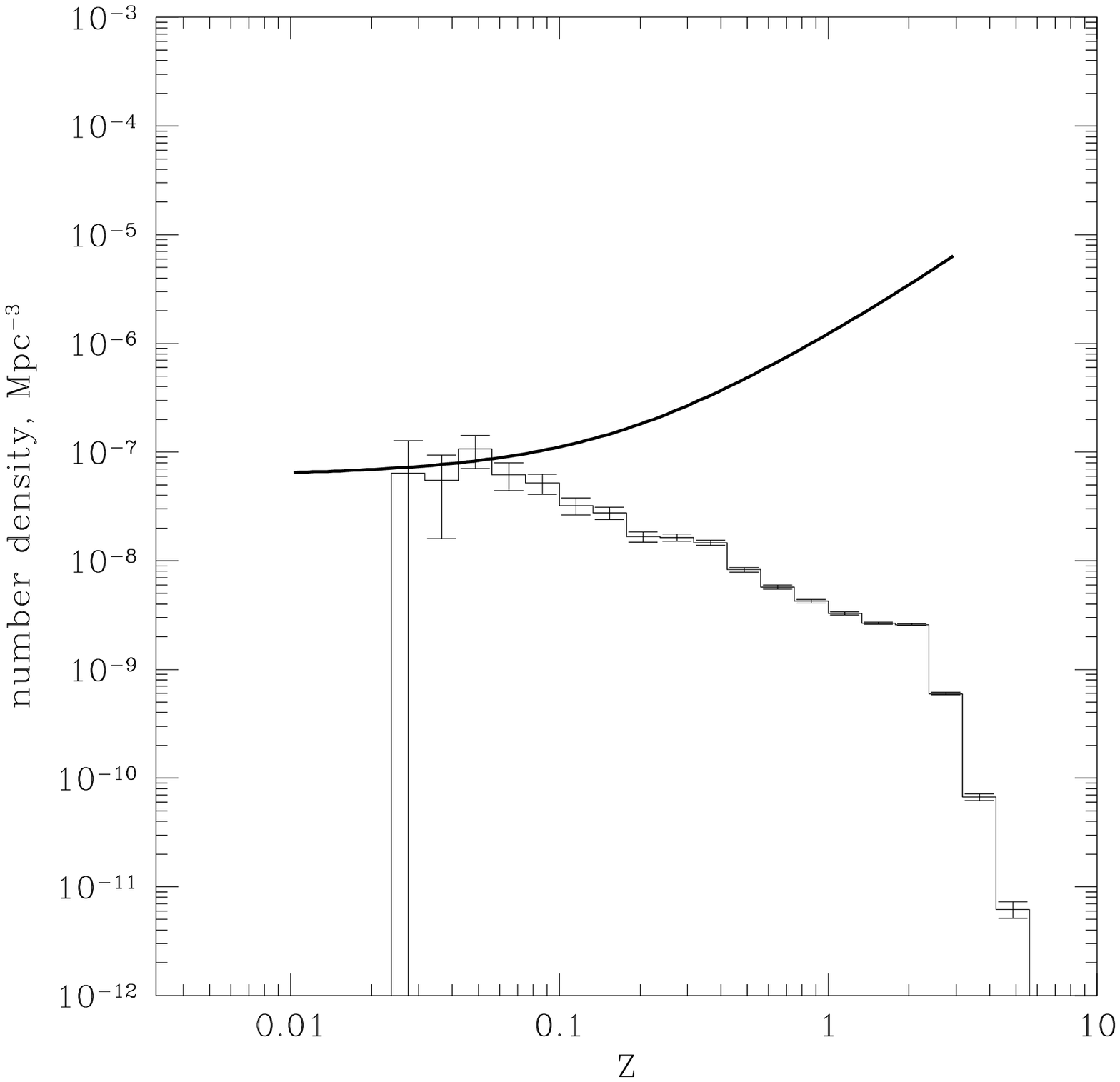}}
\vskip -5pt
\small {\bf Fig.~1}--- Number density of QSO from VCV96 as a function of
redshift. Smooth solid line represents the true number density of these
objects (accounting for their cosmological evolution, Franceschini \etal\
1994).
\end{minipage}
\hfill
\begin{minipage}[t]{0.47\textwidth}
\epsfysize=1.1\textwidth
\centerline{\epsffile{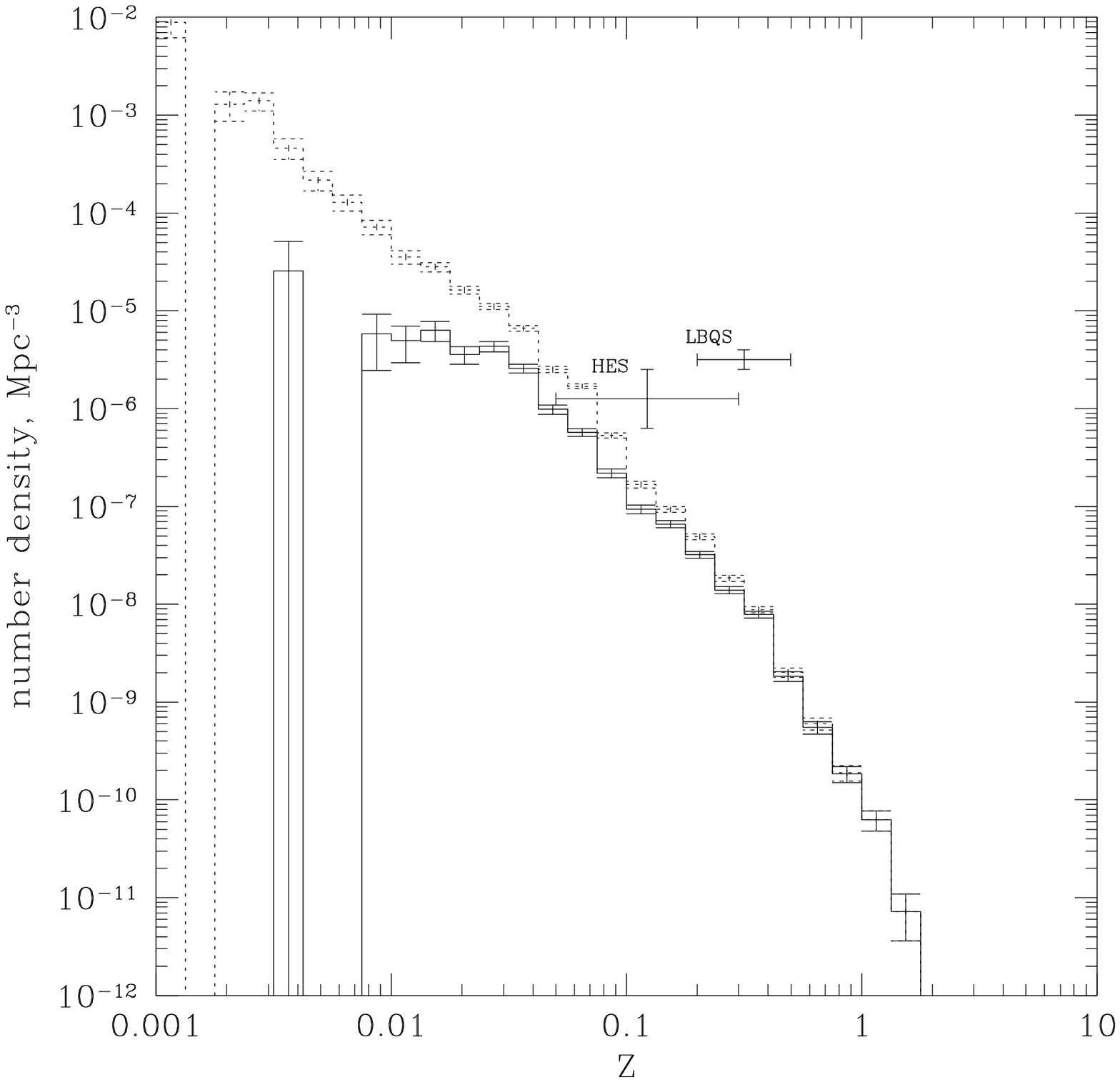}}
\vskip -5pt
\small
{\bf Fig.~2}--- Number density of AGN from VCV96 as a function of their
redshift.  Dotted histogram represents number density of all AGN,
the solid one corresponds to bright $M_B<-21$ AGN. 
\end{minipage}
\end{figure*}

\section{The data}
\subsection{Bursts localizations}

We use 327 published gamma-ray bursts with positions known to better than
$\sim 1^\circ$.  We divide all bursts in three sets basing on the origin of
the burst position measurement. The first set includes 43 burst positions
measured by WATCH/GRANAT (Sazonov et al.\ 1998), 12 --- by WATCH/EURECA
(Brandt et al.\ 1995), 30 --- by IPN 78--80 (Atteia et al.\ 1987). Thirteen
more localizations were found in the literature cited by Lund (1995). All
these sources provide 99\% confidence burst position regions. In total, the
first set contains 98 bursts.

The second and third data sets are based on BATSE data. We use only bursts
with BATSE error circle $<1.8^\circ$ at 68\% confidence. The second data set
includes 117 such bursts from the third BATSE catalog (Meegan \etal\ 1996),
excluding those GRB which are already in the first set. The third data set
includes 112 bursts from the current BATSE catalog, excluding bursts already
in the first or second sets. The current catalog was download on October 13,
1997 from http://www.batse.msfc.nasa.gov/.

Peak fluxes of the GRB in use are in the range
$\sim3-30\times10^{-6}$~erg~s$^{-1}$~cm$^{-2}$ in the 100--500 keV energy
band; the peak fluxes, and hence the burst distance distributions, are
approximately the same in all three datasets.

\subsection{Catalog of QSO and AGN}

We used QSO and AGN from V\`eron-Cetti \& V\`eron (1996; VCV96 hereafter)
``Catalogue of Quasars and Active Nuclei'' compiled from the literature.
This catalog contains 8609 QSO, 220 BL Lac objects, and 2833 AGN.  The
catalog completeness varies with the object class and redshift which makes a
secondary selection warranted.  Since we have only 327 GRB, it is
meaningless to correlate their positions with objects from a catalog with
completeness $<0.3\%$. We attempted to define subsamples of VCV96 catalog
which are complete at the level sufficient to search for angular correlation
of objects and GRB.

Figure~1 shows number density of QSO from VCV96 as a function of redshift.
The distribution predicted for the pure luminosity evolution model of
Franceschini et al.\ (1994) is also shown.  It is clear that VCV96 catalog
of QSO is almost complete at $z<0.1$; it is less than $0.3\%$ complete at
$z>1$ and it is meaningless to search for correlation of our GRB and QSO at
this redshift. Similarly, we find that the catalog becomes essentially
incomplete for BL Lac objects at $z\gtrsim0.3$ and for AGN already at
$z\sim0.03$ (Fig.~2).  As was discussed in \S1, sources of bright bursts are
probably located at higher redshifts, $z\sim0.1$--$1$, therefore the search
for correlation of GRB and all AGN from VCV96 can be inconclusive. The
subsample of intrinsically luminous AGN, $M_B<-21$, is more complete at high
redshifts.  The solid line in Fig.~2 shows the redshift distribution
$M_B<-21$ AGN.  We also show the number densities expected from AGN
luminosity functions in Hamburg/ESO (HES, K\"ohler \etal\ 1997) and the
Large Bright Quasar (LBQS, Hewett \etal\ 1993) surveys. It is clear from
Fig.~2 that VCV96 catalog is reasonably complete for $M_B<-21$ AGN out to
$z\sim 0.3$. We will search for correlation of GRB with these luminous AGN.

\section{Data analysis}

A naive method to search for correlation of objects and GRB would be to
count the number of objects inside the GRB position error boxes and compare
this number with that expected for random burst/object locations. However,
the spatial distribution of AGN and QSO from the VCV96 catalog is very
non-uniform, because many objects were found in deep, small area surveys. In
this case, it is better to use a statistics, which we denote $\delta$, of
the number of bursts whose position error regions contain at least one AGN
or QSO.  The distribution of $\delta$ expected for purely chance
associations can be derived by Monte-Carlo simulations in which the actual
AGN and QSO catalog is cross-correlated with mock catalogs of randomly
located GRB. The mock GRB catalogs were simulated with the same position
error boxes as in the data, but with centroid positions distributed
randomly. In simulating GRB positions, we accounted for the sky exposure for
WATCH/GRANAT and BATSE bursts, and assumed a uniform coverage for other
experiments. We have found that the distribution of $\delta$ derived from
simulations is accurately approximated by the binomial distribution with the
number of trials, $n$, equal to the number of bursts, and success
probability $p=\avdelta/n$, where $\avdelta$ is the average value derived
from simulations. This binomial approximation was used to estimate the
probability of strong deviations of $\delta$ found in our analysis.

\begin{table*}[tb]
\small
\centerline{{\bf Table 1} --- Correlation analysis without
redshift constraints}
\footnotesize
\medskip
\renewcommand{\arraystretch}{1.3}
\renewcommand{\tabcolsep}{0.24cm}
\begin{center}
\begin{tabular}{rcccccccc}
\hline
\hline
& & Number of bursts & $\avdelta$ & $\delta_0$ & 
$\mathrm{P}(\delta\geq\delta_0)$, \%\\
\hline
\multicolumn{1}{l}{\rlap{QSO (8609 objects):}} \\
\hline
 dataset \#1  & & 98  &  9.41 &  13 & 15.6 \\
 dataset \#2  & & 117 & 46.26 &  55 &  9.9 \\
 dataset \#3  & & 112 & 44.31 &  47 & 34.3 \\
  all         & & 327 & 96.02 & 115 &  2.6 \\
\hline
\multicolumn{1}{l}{\rlap{BL Lac (220 objects):}} \\
\hline
 dataset \#1  & & 98  &  0.66 & 2 & 14.2 \\
 dataset \#2  & & 117 &  5.50 & 2 & 97.4 \\
 dataset \#3  & & 112 &  5.04 & 2 & 96.1 \\
 all          & & 327 & 11.19 & 6 & 96.4 \\
\hline
\multicolumn{1}{l}{\rlap{Bright AGN (1390 objects):}} \\
\hline
 dataset \#1  & & 98  &  3.38 &  6 & 12.7 \\
 dataset \#2  & & 117 & 22.26 & 26 & 21.4 \\
 dataset \#3  & & 112 & 20.94 & 25 & 18.7 \\
 all          & & 327 & 45.92 & 57 &  5.1 \\
\hline
\end{tabular}
\end{center}
\end{table*}

\begin{figure*}[htb] 
\vskip 20pt
\centerline{\hfill QSO \hfill\hfill AGN\hfill}
\vskip -35pt
\centerline{\epsfysize=0.465\textwidth\epsffile{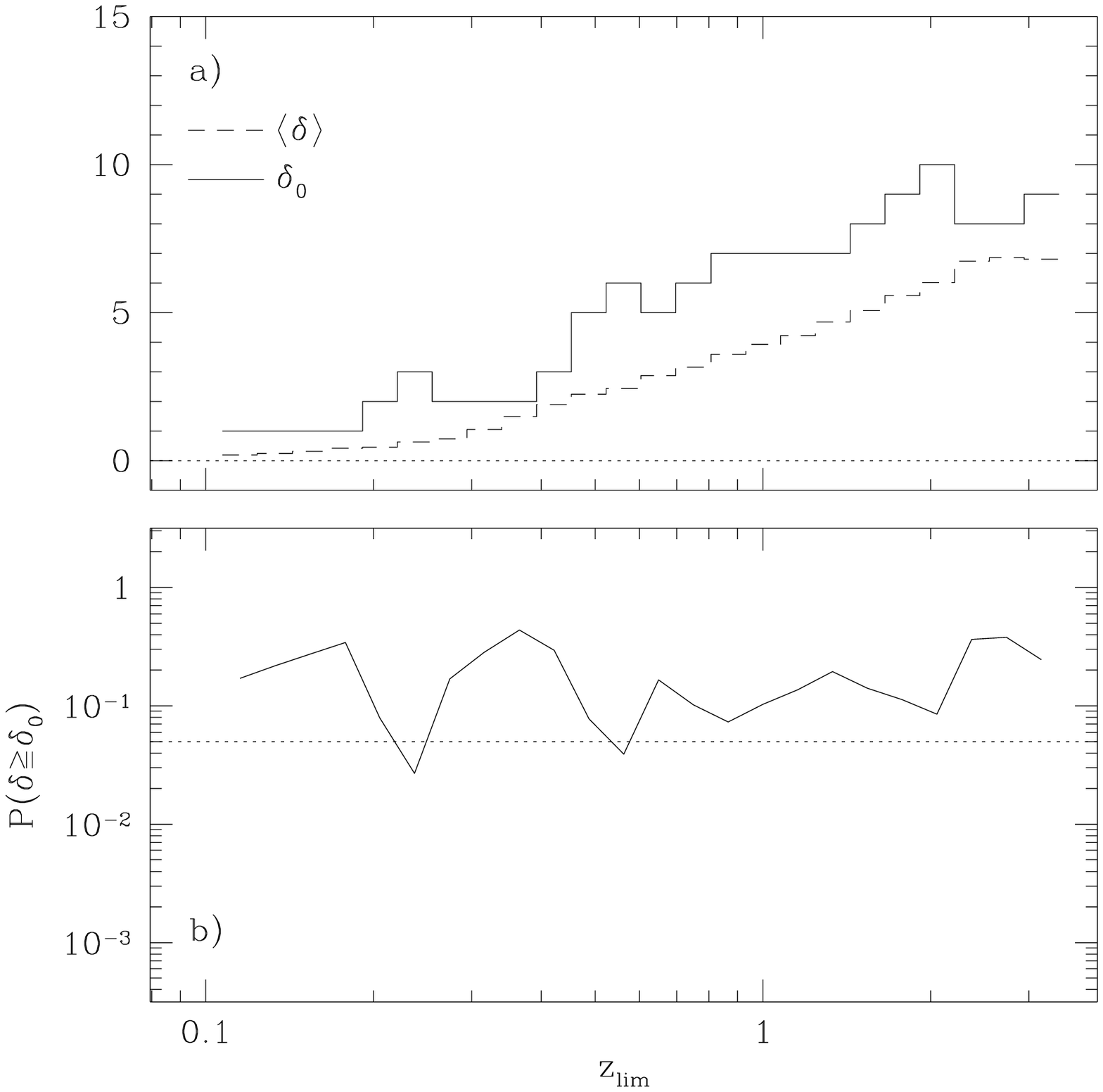} \hfill 
\epsfysize=0.465\textwidth\epsffile{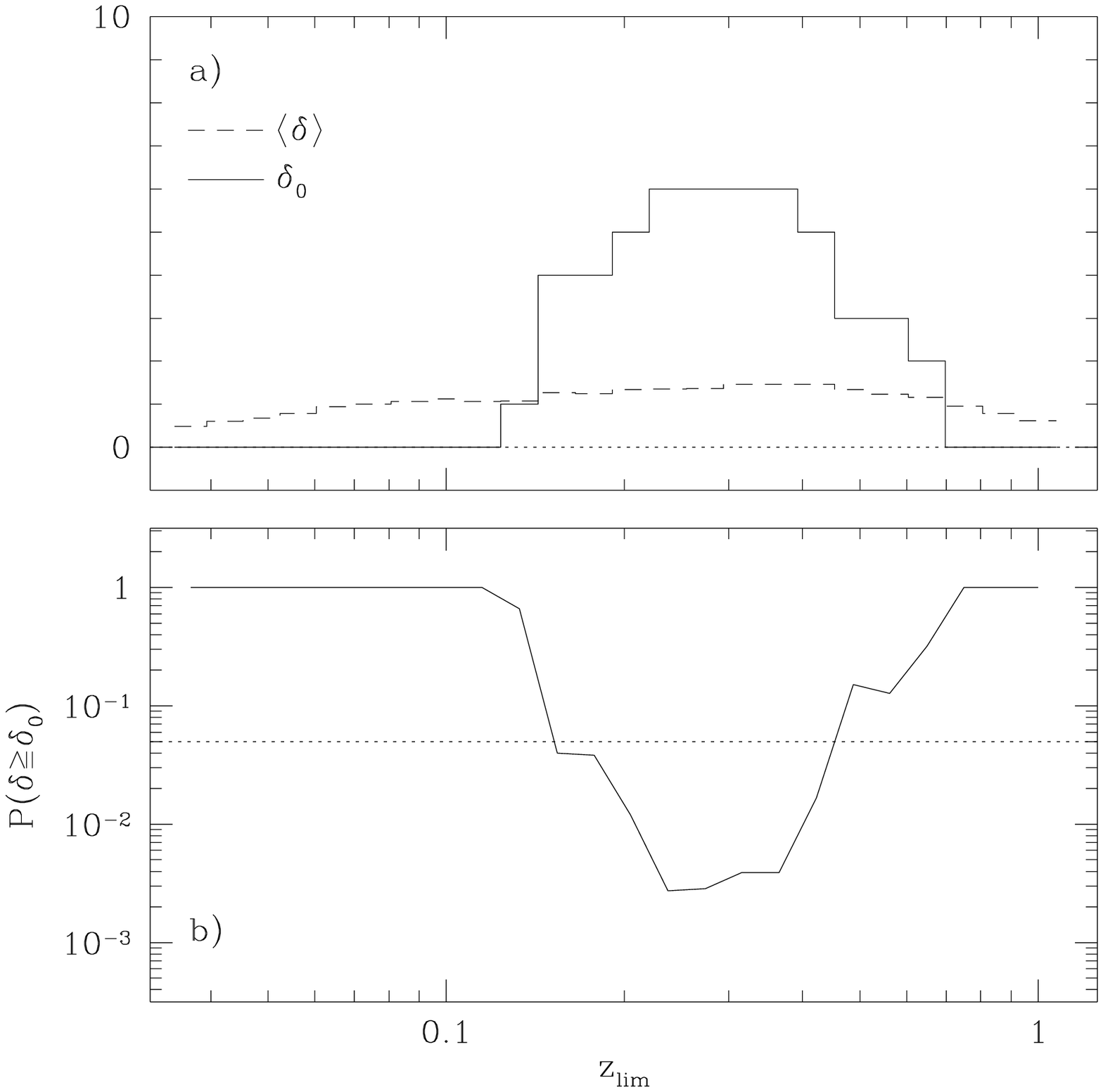}}
\vskip -7pt
\small
{\bf Fig.~3} --- Correlation analysis of the first GRB dataset with the QSO
and AGN. Solid lines in the upper panels show the values of $\delta$ (see
text) found in the data with the redshift constraint $0.32z_{\rm
lim}<z<z_{\rm lim}$ applied. The dashed lines in the upper panel show the
expected values of $\delta$ for a purely random position association between
GRB and QSO/ANGs. The lower panels show the probability of producing the
observed $\delta$'s from random fluctuations. The dotted line in the lower
panels show the 95\% ($2\sigma$) significance level. 
\end{figure*}

\begin{figure*}[htb]
\vskip 20pt
\centerline{\hfill QSO \hfill\hfill AGN \hfill}
\vskip -35pt
\centerline{\epsfysize=0.465\textwidth\epsffile{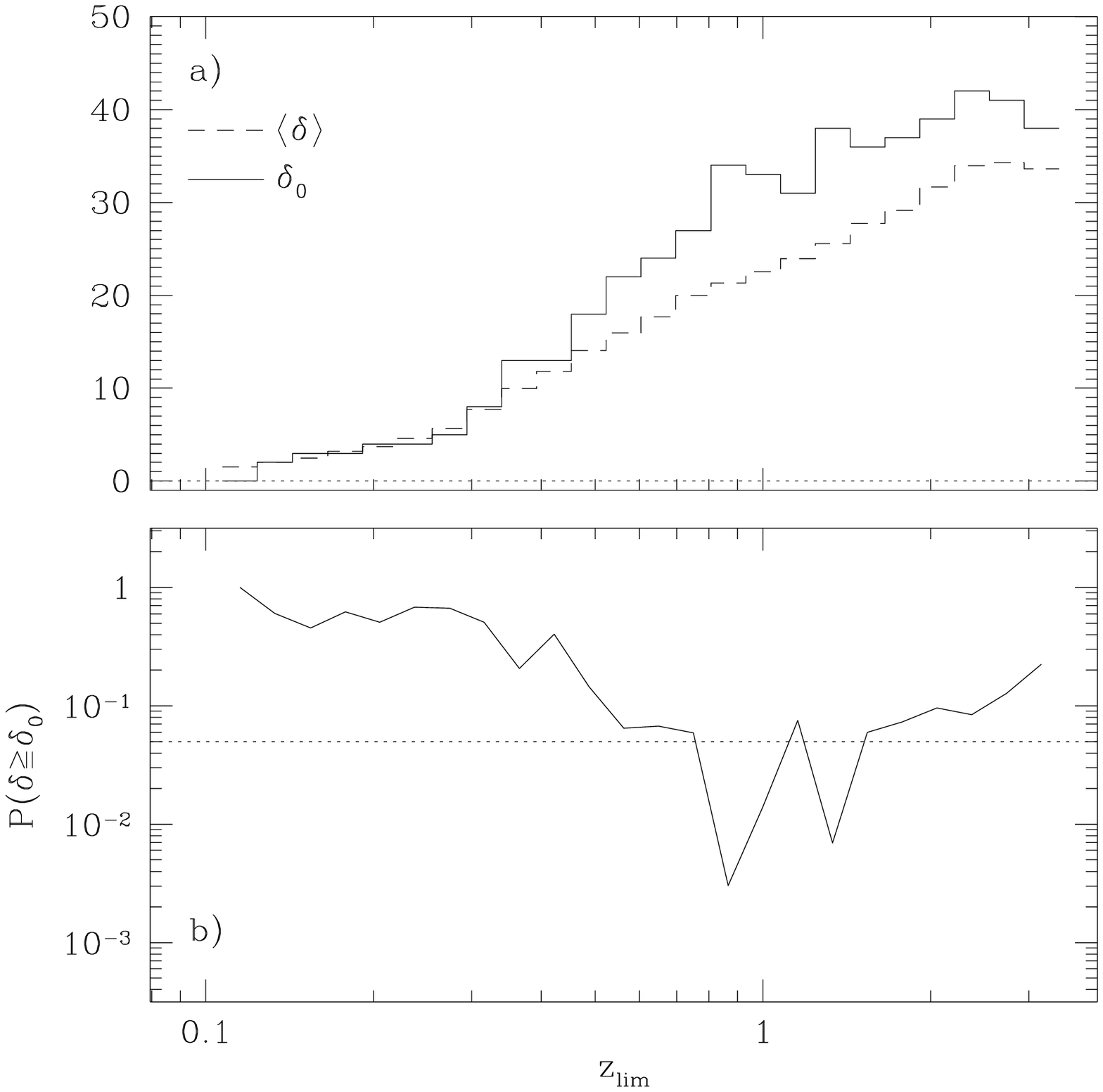} \hfill 
\epsfysize=0.465\textwidth\epsffile{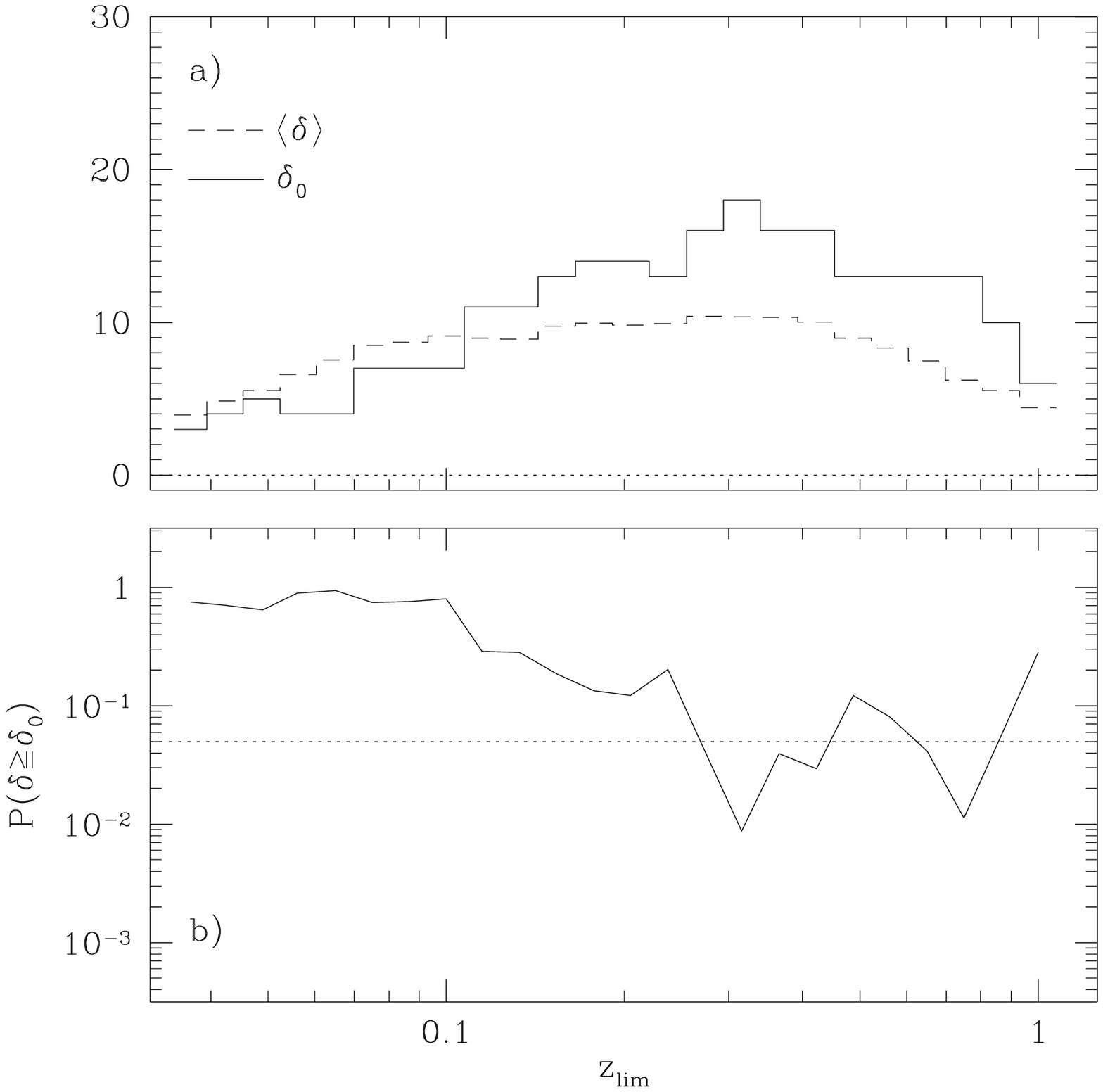}}
\vskip -7pt
\small
{\bf Fig.~4} --- Same as Fig.~3, but for the second GRB dataset (the third
BATSE catalog).
\end{figure*}

\begin{figure*}[htb] 
\vskip 20pt
\centerline{\hfill QSO \hfill\hfill AGN\hfill}
\vskip -35pt
\centerline{\epsfysize=0.465\textwidth\epsffile{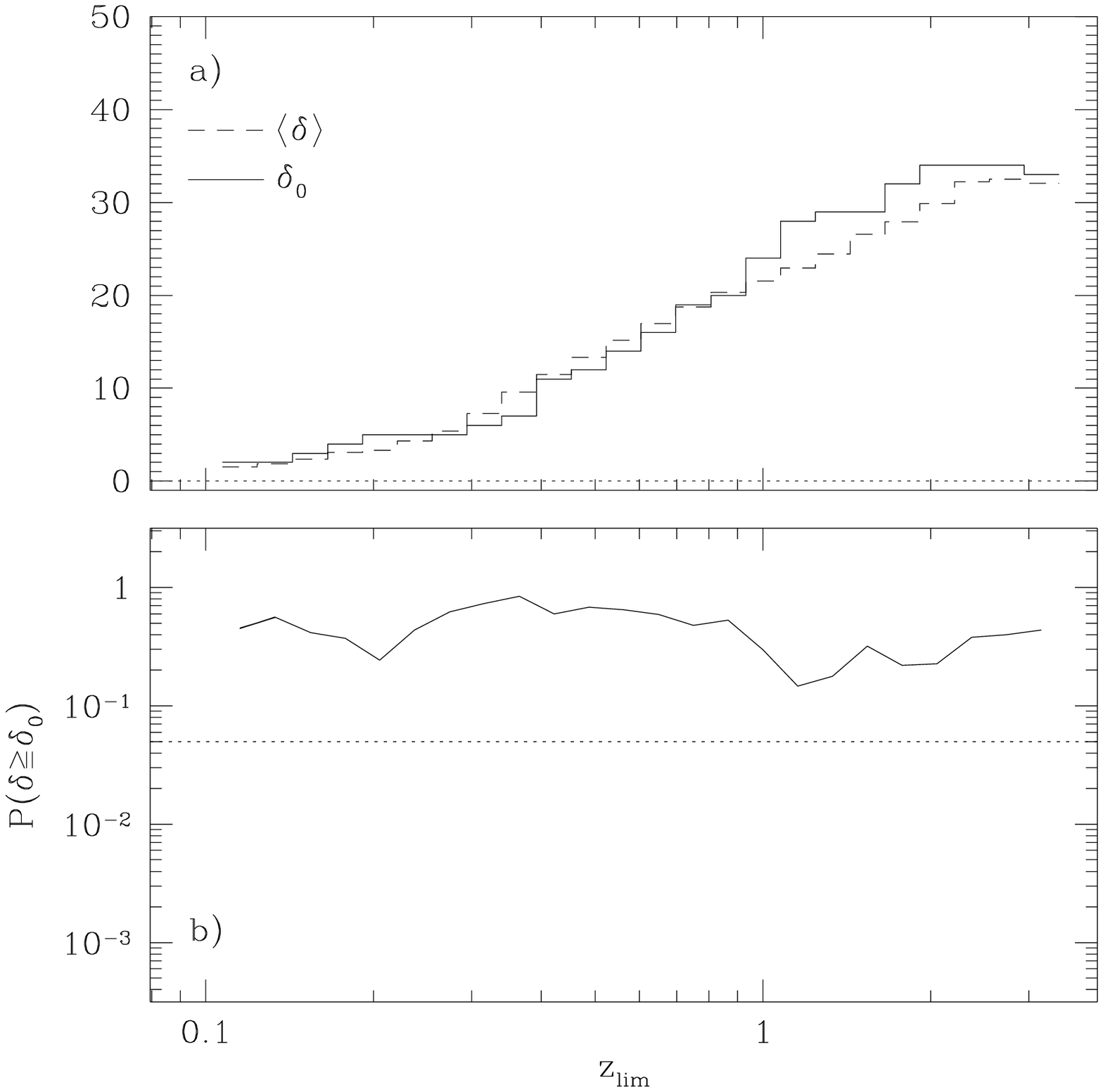} \hfill 
\epsfysize=0.465\textwidth\epsffile{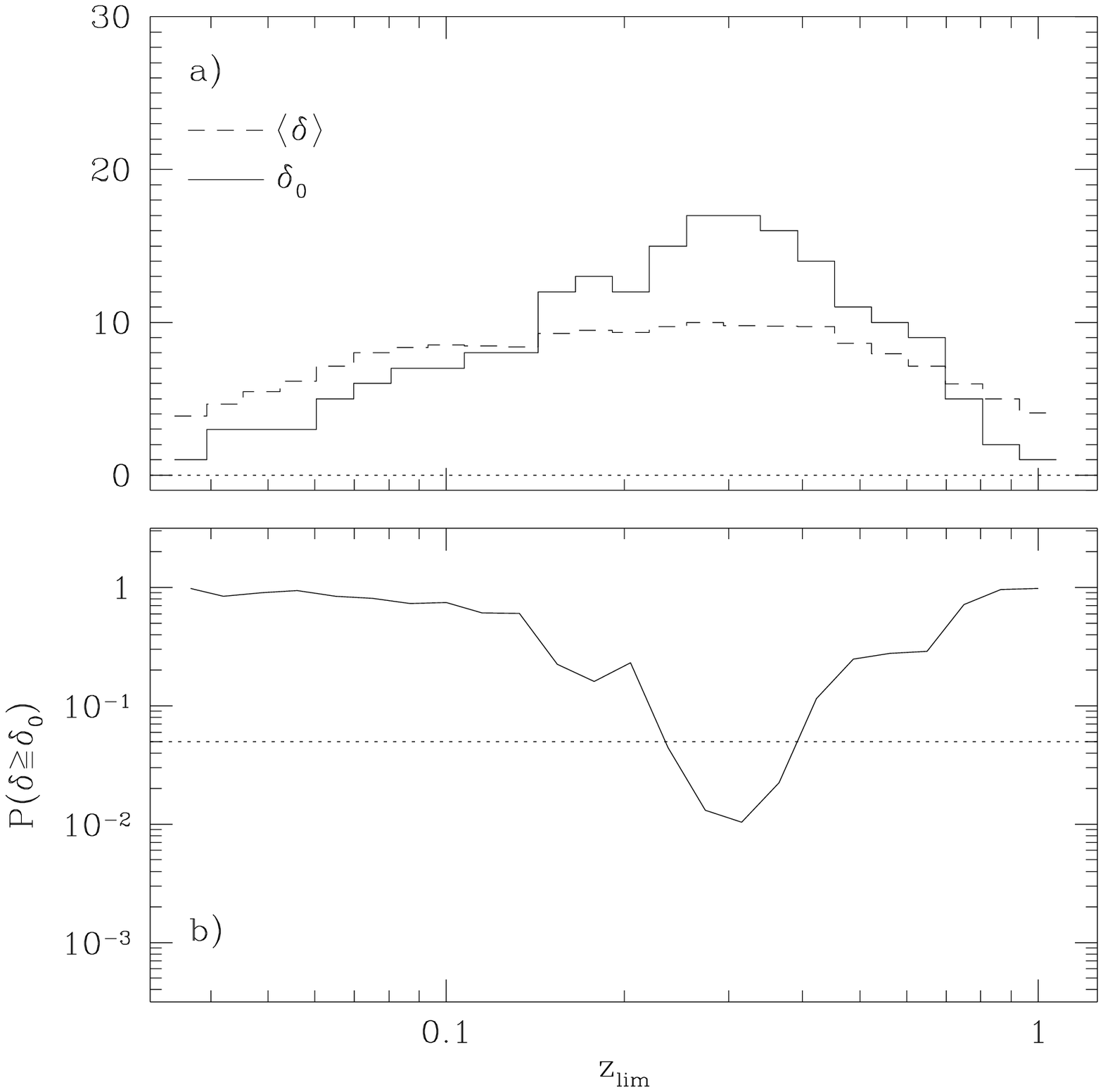}}
\vskip -7pt
\small
{\bf Fig.~5} --- Same as Fig.~3, but for the third GRB dataset (the current
BATSE catalog).
\end{figure*}

\begin{figure*}[htb] 
\vskip 20pt
\centerline{\hfill QSO\hfill\hfill AGN\hfill}
\vskip -35pt
\centerline{\epsfysize=0.465\textwidth\epsffile{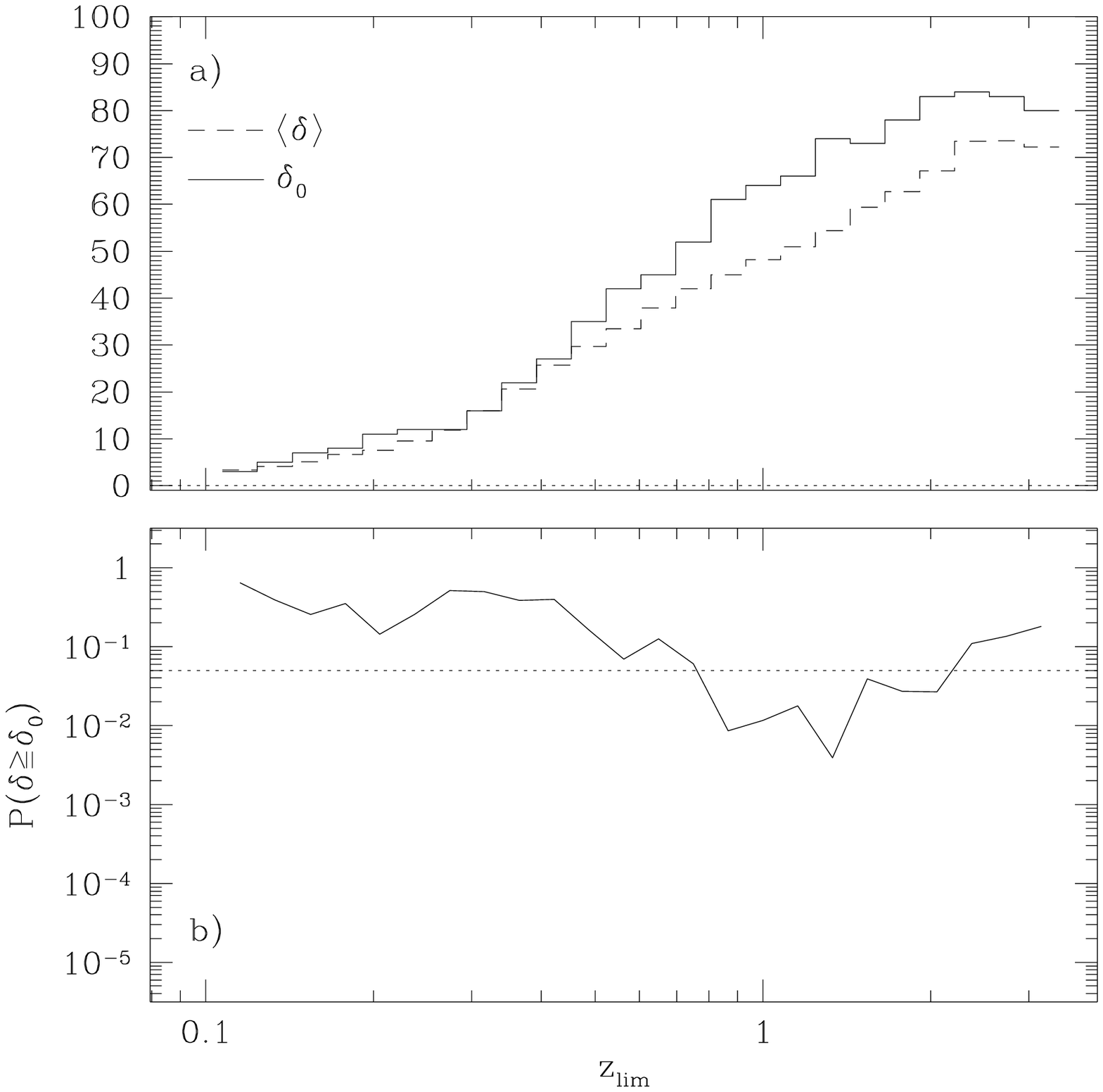} \hfill 
\epsfysize=0.465\textwidth\epsffile{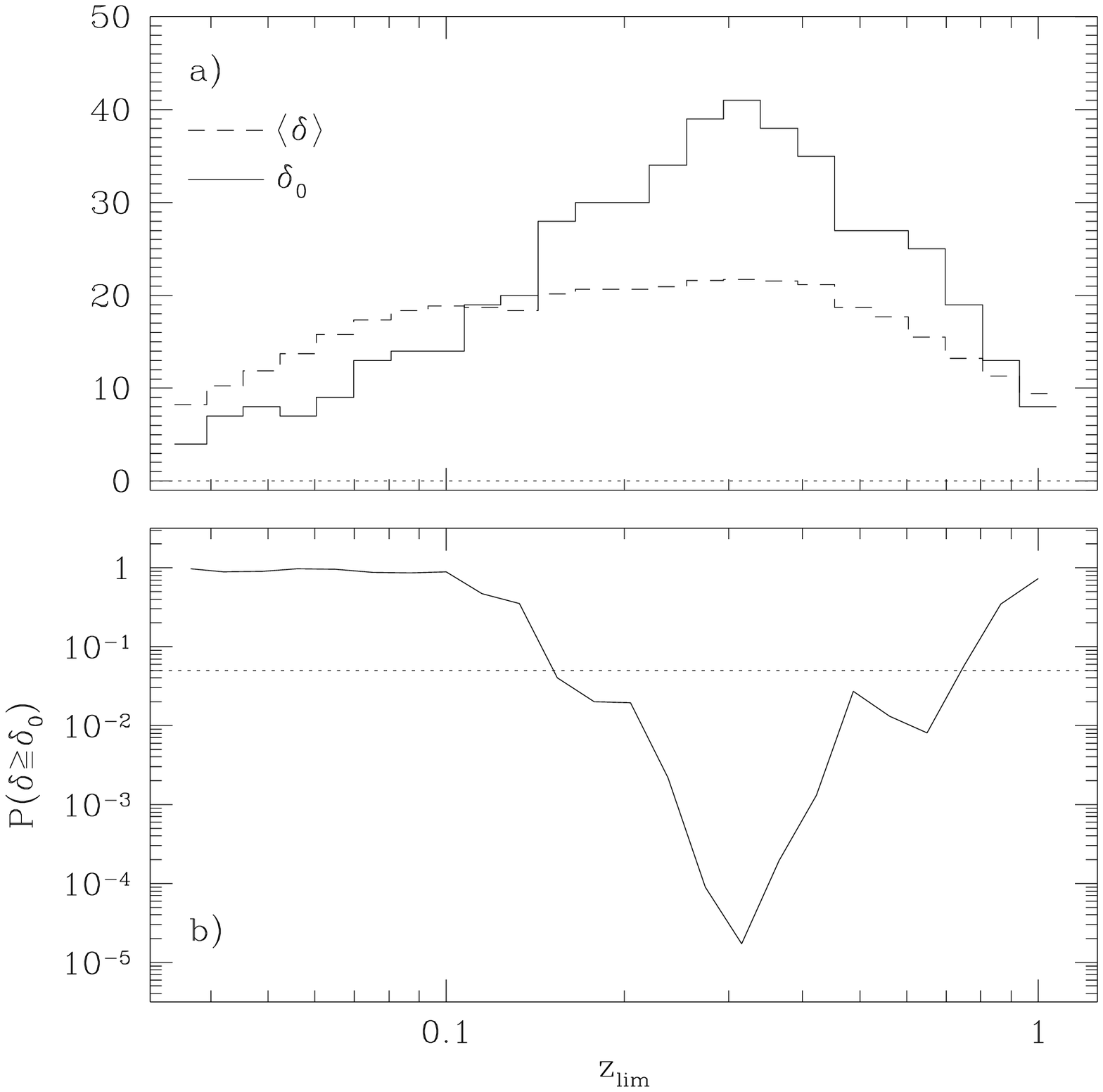}}
\vskip -5pt
\small
{\bf Fig.~6} --- Same as Fig.~3, but for all three GRB datasets (327 bursts
in total).
\end{figure*}

In search for angular correlation, it is reasonable to restrict the AGN
redshift range since there are indications that the GRB luminosity function
must be narrow, $\Delta \lg L \approx 1$ (Emslie \& Horack 1994). If this is
the case, the flux range of our bursts corresponds to the distance range of
a factor of $\lesssim 10$. Therefore, the sensitivity can be enhanced by
considering only AGN with $0.1\zlim < z < \zlim$ and varying $\zlim$. The
sensitivity may be enhanced still further if the lower boundary is a larger
fraction of $\zlim$, because the VCV96 catalog completeness falls rapidly
with redshift, while most of the volume (and presumably, most of the GRB
sources) is near the upper boundary of the redshift interval.  Therefore, we
use the redshift range $0.32\zlim < z < \zlim$. In doing so, we retain
approximately 97\% of the volume of the original $0.1\zlim < z <
\zlim$ interval and in the same time significantly improve the sensitivity.

\section{Results}

As a first step, we performed the correlation analysis for QSO, BL Lac
objects, and bright AGN without any redshift constraints. The results are
presented in Table~1, where we list the observed values of $\delta$ (recall
that $\delta$ is the number of bursts with at least one cataloged object
inside the error box), the average values $\avdelta$ expected for purely
chance associations of bursts and objects, and the probability to find a
larger value of $\delta$ than observed. In neither case, a significant (at
at least 99\% confidence) correlation is found.

Next, we performed the correlation analysis with the redshift constraints as
described in \S3 above. In particular, we considered QSO or AGN\footnote{The
number of BL Lac objects in the VCV96 catalog is too small to analyze
different redshift intervals separately, so we do not consider them below.}
in the redshift interval $0.32\zlim < z < \zlim$, and determined the values
$\delta_0$, $\avdelta$, and $\mathrm{P}(\delta>\delta_0)$ as a function of
limiting redshift $\zlim$.  Results for the first GRB dataset are shown in
Fig.~3 for both QSO and AGN.  While QSO show a slight excess of $\delta$
over random expectations, this excess is $<2\sigma$ significant at almost
all $\zlim$.  AGN, on the contrary, show a profound peak of correlation for
$\zlim\sim0.3$, i.e.\ in the redshift interval $0.1\lesssim z \lesssim0.3$.
The excess of the observed value of $\delta$ over $\avdelta$ is significant
at the 99.8\% confidence.

The results for 117 bursts from the second GRB dataset (the third BATSE
catalog), 112 bursts from the third GRB dataset (the current BATSE catalog),
and all three datasets combined (327 bursts in total) are shown in Figs.~4,
5, and 6, respectively. The peak of the GRB-AGN correlation amplitude at
$\zlim\approx0.3$ found in the first dataset, is present in two other
datasets. When all three GRB datasets are combined (Fig.~6), the statistical
significance of this correlation is very high, $99.9985\%$. Of course, the
AGN redshift range was optimized, therefore the actual significance is
somewhat lower. The number of independent redshift intervals in our analysis
can be estimated as the ratio of the log width of the searched redshift
range $0.03<z<1$ and the log width of individual intervals $\Delta\lg z
\approx 0.5$. Allowing for the partial overlap, we find $\sim 10$
independent redshift ``trials'', and hence the actual significance of the
correlation is $\sim 99.985\%$.

Gamma-ray bursts from the second dataset show some correlation with QSO
with a $\sim 99.7\%$ significance (Fig.~4). It is this correlation that was
found by Schartel \etal\ (1997). However, this correlation is absent in two
other datasets (Fig.~3 and Fig.~5); hence, the GRB-QSO correlation in Fig.~4
probably can be explained by a random deviation.

We conclude that the most prominent correlation of gamma-ray bursts with
extragalactic objects found so far is that with the Active Galactic Nuclei
in the redshift range $0.1\lesssim z \lesssim 0.3$. This correlation
suggests that at least some bursts may be physically related to AGN. Next,
we want to determine the fraction of bursts related to AGN in the redshift
interval $0.1<z<0.32$. Since not all AGN in this interval are cataloged, we
first estimated the fraction of bursts which may be related to AGN {\em from
the VCV96 catalog}. To derive the fraction of bursts which may be related to
{\em both cataloged and not cataloged\/} AGN, this fraction must be divided
by the catalog completeness.

\begin{figure*}[tb] 
\epsfxsize=0.8\textwidth
\centerline{\epsffile[103 294 520 498]{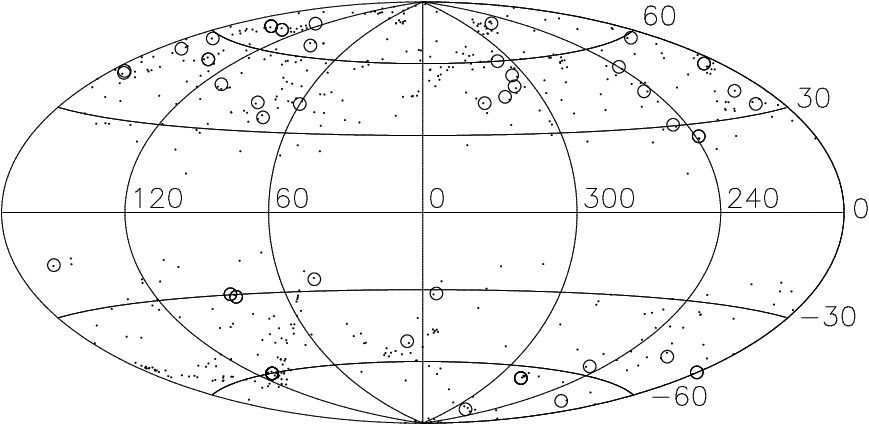}}
\bigskip
\small
{\bf Fig.~7} --- The distribution of $0.1<z<0.32$ AGN from the VCV96 catalog
in Galactic coordinates. Circles mark positions of GRB which have at least
one AGN within the position error region.
\end{figure*}

The fraction of bursts originating in cataloged AGN, $f_b$, has been
estimated as follows. Given $f_b$, the number of physically connected
bursts, $n_b$, was drawn from the binomial distribution with success
probability $f_b$ and the number of trials equal to the number of bursts. We
then randomly picked $n_b$ AGN and $n_b$ bursts from the data, and set burst
error region centroids at the chosen AGN. The positions for the rest of GRB
were chosen randomly, as in \S3. We then found the value of $\delta$ for
such a mock burst catalog. Repeating these simulations, we derived the
probability distribution $P(\delta)$ as a function of $f_b$. Using the
derived $P(\delta, f_b)$ we estimated the best-fit $f_b$ as a value which
maximized the probability to obtain the observed value $\delta_0$, i.e.\
which miminized $C=-2\log L = -2 \log P(\delta_0, f_b)$.  Here $C$ is
equivalent to the C-statistics of Cash (1979), and so the confidence
intervals of $f_b$ can be derived from the change in $C$.

This analysis was performed separately for BATSE (second+third datasets) and
non-BATSE (first dataset) burst data. In the case of BATSE data with 68\%
position error circles, the derived value $f_b$ has been corrected by a
factor of 1.47. For both BATSE and non-BATSE data, we obtained consistent
fractions, of order of several percent, of bursts originating in the
cataloged AGN. For the first dataset, $f_{b}=0.040^{+0.025}_{-0.023}$
(68\%) and $^{+0.055}_{-0.036}$ (95\% confidence). For the bursts detected
by BATSE (second and third data sets together):
$f_{b}=0.098^{+0.037}_{-0.038}$ (68\%) and $^{+0.080}_{-0.070}$ (95\%).

Next, we have to correct $f_b$ for the incompleteness of the VCV96 catalog.
If one assumes that at $z<0.03$, all $M_B<-21$ AGN are cataloged, the
completeness in the $0.1<z<0.32$ interval is low, $0.007$.
%
%
If, instead, the HES survey luminosity function is used, the completeness is
$\sim 0.03$. Both these values are smaller than the estimated $f_b$, which
means, formally, that more than 100\% of bursts originate in AGN. Clearly,
this is unphysical, and we discuss some possible explanations below.
Nevertheless, similar values of $f_b$ and catalog completeness suggest that
a very significant fraction of GRB is related to AGN. Below we discuss this
result in connection with previous attempts to identify optical counterparts
of gamma-ray bursts.

\section{Discussion}

This is not the first attempt to find optical counterparts of GRB by
analyzing the optical content of small-area localizations, so the comparison
of our results with some previous works is in order.

Vrba et al.\ (1995) have performed extensive optical photometry of 8
small-area ($<70$~arcmin$^2$) IPN localizations searching for objects with
unusual colors, variability, and proper motions. Only blue objects, which
Vrba et al.\ interpreted as QSO, showed a marginal excess at a rate
approximately one per localization. This is in agreement with our results.

Webber et al.\ (1995) and Gorosabel et al.\ (1995) used several tens GRB
localizations from IPN and WATCH, respectively. The optical content was
taken from existing catalogs, similar to our study. No excess of either AGN
or any other class of objects was found. These results do not directly
contradict to ours because Webber et al.\ (1995) and Gorosabel et al.\
(1995) used a smaller number of bursts, did not account for the
incompleteness of optical catalogs (\S2.2), and did not apply redshift
constraints in search for correlation. Since the completeness of the VCV96
catalog (and probably other all-sky catalogs) is below, or on the level of,
several percent (\S4), analyses based on several tens of bursts are
inconclusive.

The most similar to ours is the work of Schartel et al.\ (1997) who
cross-correlated BATSE bursts with the AGN and QSO from VCV96. They found a
marginal correlation with QSO, and no correlation with AGN, with no redshift
constraints applied to AGN. We essentially reproduce these results (Fig.~4
and Table~1). A significant correlation with AGN found in our analysis is
not found by Schartel et al.\ because they used a smaller number of bursts,
with poorer localizations (only BATSE data), and used the entire VCV96
catalog which is very incomplete for AGN even at low redshift (\S2.2).

Finally, we mention the two GRB with small-area localizations, in which AGN
were found. Drinkwater et al.\ (1997) report that the possible X-ray
counterpart of GRB~920501 is associated with a Seyfert~1 galaxy at
$z=0.315$. The first X-ray localization of a gamma-ray burst by BeppoSAX
(3$^\prime$ radius) contains a $z=1.038$ QSO (Piro et al.\ 1998). On the
other hand, optical transients associated with other BeppoSAX bursts,
GRB~970508, 970228, and 971214 probably are not AGN.

Our analysis supersedes most of earlier searches for GRB-AGN association
because we used a large GRB dataset, carefully accounted for incompleteness
of the optical catalogs, and introduced sensible object redshift
constraints. These advantages made it possible to find the strongest ever
evidence for association of GRB with a known class of extragalactic objects.
However, there are several problems with our analysis. First, we used
relatively large area gamma-ray burst localizations, and therefore had to
perform a statistical correlation rather than an object-by-object
identification.  Second, we used a sparse catalog of AGN and QSO, in which
many objects were found in a number of small-area high-sensitivity searches.
In fact, this introduces a possibility that the correlation we detect is not
with AGN but rather with objects around which the AGN we searched, e.g.\
normal galaxies or other QSO (data from Arp 1980 and Monk et al.\ 1988),
targets of {\em Einstein}\/ pointings (EMSS AGN, Stocke et al.\ 1991).
However, the sky distribution of GRB with $0.1<z<0.32$ AGN inside the
localization area, does not generally follow the regions of deep AGN surveys
(Fig.~7). Also, a clear redshift dependence of the correlation is not easily
explainable in such a scenario.

Another problem is that the amplitude of the detected correlation implies
that the fraction of bursts related to AGN is somewhat higher than the
estimated completeness of the optical catalog (\S4). We can offer two
possible explanations. First is that GRB prefer luminous AGN, for which the
catalog completeness is higher. Second is that GRB prefer a certain type of
AGN (Seyfert 1 or 2, X-ray loud or quiet, radio loud or quiet, etc.) which
is more commonly present in the VCV96 catalog than the ``average'' AGN.

Despite these problems, we believe that the case for an association of
bright gamma-ray bursts with AGN at moderate redshift is compelling.  This
case can be further proved, or disproved, by an extensive optical,
preferably spectroscopic, survey of small-area localizations of bright GRB,
similar to the work of Vrba et al.\ (1995) but using a larger number of
bursts. Another approach would be to use a more complete catalog of AGN
covering a significant fraction of the sky; unfortunately, this seems
impractical until the Sloan Digitized Sky Survey is completed.

\end{multicols}

\clearpage

We provide a list of GRB with AGN within localizations. The table lists the
experiment in which the burst was localized, burst ID, and names,
coordinates (J2000), redshifts and absolute magnitudes of matching AGN.

Note, that due to good localizations of WATCH and IPN bursts, we expect that
$\sim 4$ out of 7 AGN are truly related to bursts (see Fig.~3). For BATSE
bursts, which have poorer localizations, we expect that $\sim 15$ out of 53
AGN are associated with bursts (Fig.~4 and 5).

\footnotesize
\vfill

\begin{center}

{\bf Table 2} --- Gamma-ray bursts with AGN within localizations
\medskip

\def\arraystretch{1.2}
\begin{tabular}{lllcccc}
\hline
\hline
Experiment & Burst ID & AGN Name & RA  & Dec  & $z$ & $M_B$ \\
\hline
W/GR &910627&  Q 1313$-$0153     & 13 16 24.3& $-$02 09 45 & 0.150&  $-$21.6\\
W/GR &910927&  0313$-$428        & 03 14 55.9& $-$42 40 56 & 0.126&  $-$21.4\\
W/GR &920210&  MS 10185+4830     & 10 21 38.8&   +48 15 10 & 0.232&  $-$22.1\\
     &      &  Q 1019+4750       & 10 22 57.0&   +47 34 55 & 0.144&  $-$21.2\\
W/GR &920720&  HE 0936$-$1058    & 09 39 11.4& $-$11 11 46 & 0.214&  $-$22.4\\
W/GR &940703&  MS 08498+2820     & 08 52 48.8&   +28 08 40 & 0.197&  $-$21.9\\
IPN  &790101&  Q 1215+1545       & 12 17 32.8&   +15 28 45 & 0.139&  $-$21.7\\
BATSE&910507&  PKS 2004$-$447    & 20 07 55.1& $-$44 34 43 & 0.238&  $-$21.6\\
BATSE&911126&  Q 1032+062        & 10 35 06.0&   +06 01 41 & 0.245&  $-$22.2\\
BATSE&920110&  HE 0348$-$2226    & 03 50 19.2& $-$22 17 22 & 0.111&  $-$22.0\\
BATSE&920221&  Q 1206+4748       & 12 09 05.9&   +47 32 07 & 0.182&  $-$21.3\\
     &      &  Q 1221+4752       & 12 23 47.6&   +47 35 35 & 0.307&  $-$21.5\\
BATSE&920517&  IRAS 13305$-$1739 & 13 33 16.5& $-$17 55 00 & 0.148&  $-$21.2\\
BATSE&920622&  PC 1044+4719      & 10 47 13.2&   +47 03 35 & 0.247&  $-$21.5\\
BATSE&921206&  Q 1138+4638       & 11 40 48.2&   +46 22 03 & 0.115&  $-$22.0\\
     &      &  RX J11427+4625    & 11 42 41.4&   +46 24 36 & 0.114&  $-$22.0\\
     &      &  Q 1145+4638       & 11 47 47.6&   +46 21 46 & 0.151&  $-$21.1\\
BATSE&921209&  SBS 1116+610      & 11 19 22.3&   +60 48 51 & 0.298&  $-$22.9\\
     &      &  SBS 1121+606      & 11 24 17.9&   +60 20 26 & 0.200&  $-$22.5\\
BATSE&930106&  Q 0015+0119       & 00 17 45.9&   +01 36 20 & 0.236&  $-$22.5\\
     &      &  PB 5853           & 00 18 22.1&   +01 19 01 & 0.160&  $-$22.8\\
     &      &  Q 0017+0212       & 00 20 33.2&   +02 28 52 & 0.256&  $-$22.5\\
     &      &  UM 228B           & 00 21 02.3&   +00 52 41 & 0.142&  $-$22.7\\
     &      &  Q 0019+0022A      & 00 21 41.0&   +00 38 41 & 0.314&  $-$22.9\\
     &      &  Q 0023+0058       & 00 26 20.8&   +01 15 17 & 0.274&  $-$22.5\\
     &      &  Q 0023+0228       & 00 26 21.9&   +02 44 42 & 0.236&  $-$22.4\\
BATSE&930405&  MS 13061$-$0115   & 13 08 42.8& $-$01 31 24 & 0.111&  $-$21.1\\
BATSE&930425&  0110$-$361        & 01 12 32.7& $-$35 55 35 & 0.290&  $-$22.0\\
BATSE&930905&  MS 20395$-$0107   & 20 42 06.5& $-$00 56 57 & 0.142&  $-$21.1\\
BATSE&931024&  E 0906$-$091      & 09 08 51.2& $-$09 18 51 & 0.129&  $-$21.9\\
     &      &  E 0907$-$091      & 09 09 36.2& $-$09 18 19 & 0.253&  $-$22.9\\
BATSE&931204&  1631.9+3719       & 16 33 38.3&   +37 13 14 & 0.115&  $-$21.2\\
BATSE&940129&  MS 08303+2828     & 08 33 25.9&   +28 17 53 & 0.283&  $-$21.5\\
BATSE&940704&  Q 1400+4638       & 14 02 08.7&   +46 24 13 & 0.236&  $-$21.8\\
BATSE&940806&  0146$-$502        & 01 48 19.5& $-$50 02 59 & 0.310&  $-$22.9\\
     &      &  0147$-$511        & 01 49 04.4& $-$50 53 06 & 0.170&  $-$21.1\\
     &      &  0149$-$505        & 01 51 18.5& $-$50 16 26 & 0.310&  $-$22.8\\
     &      &  0149$-$510        & 01 51 44.6& $-$50 50 08 & 0.290&  $-$22.2\\
BATSE&940830&  MS 22229+2046     & 22 25 19.3&   +21 01 45 & 0.139&  $-$22.3\\
     &      &  MS 22230+2110     & 22 25 23.9&   +21 25 25 & 0.310&  $-$22.9\\
\hline
\end{tabular}
\end{center}

\vfill
\mbox{}

\clearpage
\mbox{}
\vfill

\footnotesize
\begin{center}

{\bf Table 2} --- {\em Continued}
\medskip

\def\arraystretch{1.2}
\begin{tabular}{lllcccc}
\hline
\hline
Experiment & Burst ID & AGN Name & RA  & Dec & $z$ & $M_B$ \\
\hline
BATSE&950305&  RX J13113$-$1411  & 13 11 18.3& $-$14 11 28 & 0.153&  $-$21.4\\
BATSE&950904&  MS 17096+4823     & 17 10 58.5&   +48 19 27 & 0.174&  $-$22.3\\
BATSE&951002&  PB 6986           & 03 02 46.8& $-$02 34 35 & 0.249&  $-$22.7\\
     &      &  Q 0307$-$0200     & 03 10 19.1& $-$01 48 40 & 0.207&  $-$22.2\\
BATSE&951016&  PKS 0222$-$23     & 02 25 02.6& $-$23 12 49 & 0.230&  $-$22.3\\
BATSE&951102&  PG 0906+48        & 09 10 09.9&   +48 13 42 & 0.118&  $-$22.8\\
BATSE&951202&  A09.64            & 21 56 03.0& $-$36 15 15 & 0.115&  $-$22.2\\
BATSE&960111&  PG 0906+48        & 09 10 09.9&   +48 13 42 & 0.118&  $-$22.8\\
BATSE&960206&  HE 1254$-$0934    & 12 56 56.9& $-$09 50 16 & 0.139&  $-$22.2\\
BATSE&960605&  Q 1010$-$0056     & 10 13 17.2& $-$01 10 57 & 0.202&  $-$22.2\\
BATSE&960804&  RX J14386+6910    & 14 38 40.7&   +69 10 40 & 0.214&  $-$22.8\\
BATSE&960807&  2E 1028+3102      & 10 31 39.0&   +30 46 46 & 0.250&  $-$21.5\\
BATSE&961202&  MS 22229+2046     & 22 25 19.3&   +21 01 45 & 0.139&  $-$22.3\\
     &      &  MS 22230+2110     & 22 25 23.9&   +21 25 25 & 0.310&  $-$22.9\\
BATSE&961213&  HS 1234+4610      & 12 36 58.0&   +45 53 52 & 0.240&  $-$22.2\\
BATSE&970223&  NGC 2859 U1       & 09 24 34.8&   +34 40 33 & 0.230&  $-$21.6\\
     &      &  MS 09227+3420     & 09 25 46.0&   +34 07 46 & 0.158&  $-$22.2\\
BATSE&970411&  3C  93.1          & 03 48 46.9&   +33 53 16 & 0.244&  $-$21.9\\
BATSE&970420&  RX J14136$-$1538  & 14 13 40.3& $-$15 38 33 & 0.226&  $-$22.9\\
BATSE&971006&  2E 1640+5345      & 16 42 00.8&   +53 39 51 & 0.140&  $-$21.6\\
\hline
\end{tabular}
\end{center}

\vfill
\mbox{}

\end{document}